\documentclass[usegraphicx]{mn2e}

\usepackage{graphicx}
\usepackage{afterpage}

\newcommand\Msun{{\,M_\odot}}

\title[]{The effect of mass ratio on the morphology and time-scales of disc galaxy mergers}

\author[J.M. Lotz et al.]{Jennifer M. Lotz,$^1$\thanks{NOAO Leo Goldberg Fellow},  
Patrik Jonsson,$^2$ T.J. Cox,$^3$\thanks{W.M. Keck Fellow} and Joel R. Primack$^2$ \\
$^1$ National Optical Astronomical Observatory, 950 N.  Cherry Avenue, Tucson, AZ 85719 USA; lotz@noao.edu \\
$^2$ Department of Physics, University of California, Santa Cruz, CA, 95064 USA \\
$^3$ Harvard-Smithsonian Center for Astrophysics, 60 Garden St.,  Cambridge, MA, 02138 USA}

\begin{document}

\date{submitted 17 August 2009; resubmitted 8 December 2009}

\pagerange{\pageref{firstpage}--\pageref{lastpage}} \pubyear{2009}

\maketitle

\label{firstpage}

\begin{abstract}
The majority of galaxy mergers are expected to be minor mergers.  
The observational signatures of minor mergers are not well understood, thus there exist
few constraints on the minor merger rate. This paper seeks to address this gap in our understanding
by determining if and when minor mergers exhibit disturbed morphologies and how they differ from the morphology of major mergers. 
We simulate a series of unequal-mass moderate gas-fraction disc galaxy mergers.  
With the resulting $g$-band images, we determine how the time-scale for identifying galaxy mergers 
via projected separation and quantitative morphology (the Gini coefficient $G$, asymmetry $A$, and the second-order moment of the
brightest 20\% of the light $M_{20}$) depends on the merger mass ratio, relative 
orientations and orbital parameters.  We find that $G-M_{20}$ is as sensitive to 9:1 baryonic mass ratio mergers as
1:1 mergers, with observability time-scales $\sim$ 0.2$-$0.4 Gyr. 
In contrast, asymmetry finds mergers with baryonic mass ratios between 4:1 and 1:1 (assuming local disc galaxy gas-fractions). 
Asymmetry time-scales for moderate gas-fraction major disc mergers are $\sim$ 0.2$-$0.4 Gyr, and less than 0.06 Gyr for
moderate gas-fraction minor mergers.  
The relative orientations and orbits have little effect on the time-scales for morphological disturbances. 
Observational studies of close pairs often select major mergers by choosing paired galaxies with similar luminosities and/or  
stellar masses.  Therefore, the various ways of finding galaxy mergers ($G-M_{20}$, $A$,  close pairs) 
are sensitive to galaxy mergers of different mass ratios. By comparing the frequency of mergers selected by different techniques,  
one may place empirical constraints on the major and minor galaxy merger rates. 
\end{abstract}


\section{INTRODUCTION}
The vast majority of galaxy mergers are expected to be mergers between unequal-mass systems.  Within the
standard cosmological framework, the frequency of dark matter halo mergers increases with the mass ratio of
the merger (e.g. Lacey \& Cole 1993; Fakhouri \& Ma 2008; Genel et al. 2009).  `Minor' mergers of galaxies with total mass ratios greater than
3:1 $-$ 4:1 are more common than `major' mergers with mass ratios closer to 1:1.  Cosmological-scale
simulations and semi-analytic models suggest that minor mergers play a significant role in triggering star-formation in the early universe
(e.g. Somerville, Primack, \& Faber 2001),  accreting mass onto galaxies (e.g. Guo \& White 2008; Genel et al. 2009; Stewart et al. 2009), 
and growing the stellar haloes of massive galaxies (e.g. Johnston, Hernquist, \& Bolte 1996; Bell et al. 2008).   

However, there exist few observational constraints on the minor merger rate and its evolution with time.  Galaxy mergers
may be identified either by searching for pre-merger systems of galaxies close in projected separation and relative velocity, or
for galaxies with disturbed morphologies.   Minor merger studies with close pairs are hampered by the high contrast ratio between
the primary and satellite galaxies.   It is challenging to obtain spectroscopic velocities or reliable photometric redshifts 
for the satellite galaxies with masses and luminosities $\sim$ a tenth of the primary galaxies.  Additionally,  the contamination rate from background
galaxies increases dramatically as the luminosity difference between the primary and satellite galaxies increases.  

\begin{figure*}
\includegraphics[width=184mm]{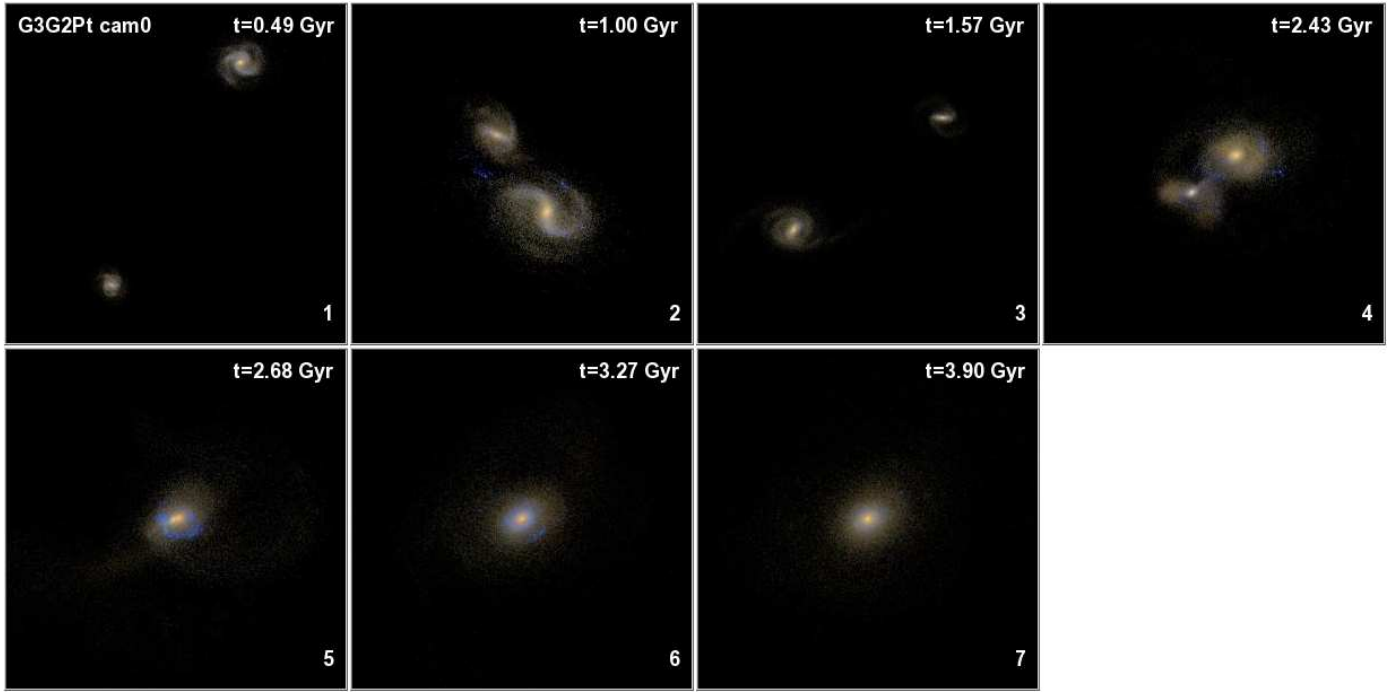}
\caption{ $u-r-z$ composite colour images including dust extinction 
for the 3:1 baryonic mass ratio  tilted prograde-prograde G3G2P$_t$ simulation.  The G3 primary galaxy is viewed face-on (camera 0).   
The images show  the initial galaxies [1], the first pass [2], the maximal separation after the first pass [3], the second
pass [4],  the final merger [5], the post-merger $\sim$ 0.5 Gyr after the merger [6], and the remnant $\sim$ 1 Gyr after the merger [7]. 
The field of view for panels 1 and 3 is 200 kpc, while the field of view for the other images is 100 kpc.   } \label{g3g2col}
\end{figure*}

It has generally been assumed that the morphological effects of minor mergers are subtle.  High resolution N-body simulations
of gas-rich minor mergers find that such events may thicken but not destroy the discs of the primary galaxies (Moster et al. 2009; 
Purcell, Kanzantzidis, \& Bullock 2009; Robertson et al. 2006; Abadi et al. 2003),  as may minor mergers on retrograde or circular orbits 
(e.g. Velazquesz \& White 1999; Hopkins et al. 2009a). 
Similarly, minor mergers may deposit material onto the outer regions of high redshift compact spheroids, but do not significantly 
change their inner density structure  (Naab et al. 2009; Hopkins et al. 2009b).  A study of a 5:1 pure N-body galaxy merger simulation found
that the merger never showed asymmetries strong enough to meet the standard merger criterion (Conselice 2006).  
However, recent studies of the visual classification of
galaxy mergers suggests that $\sim$ 50\% of visually disturbed galaxies may result from minor mergers  (Jogee et al 2009; Darg
et al. 2009).  Minor mergers and interactions may also produce lopsided discs (e.g. Rudnick \& Rix 1998; Bournaud et al. 2005;  Reichard et al. 2008).    
Therefore the morphological effects of minor mergers during the merger event may be more significant than previously thought. 

 With a better physical understanding of why and when merging galaxies 
appear morphologically disturbed, we can constrain the observed galaxy minor and major merger rate. 
 Most previous theoretical studies 
of minor mergers have focused on the structures of the merger remnants (e.g. Hopkins et al. 2009; Naab et al. 2009; Robertson et al 2006;  
Bournaud, Jog, \& Combes et al. 2004;  Johnston et al. 1996)  or spanned a very limited range of parameter space (Conselice 2006).  
Bournaud et al. (2005) performed a detailed study of the lopsidedness of disc galaxy mergers, and found that gas-rich minor mergers
could appear strongly lopsided.  In this work, we explore the strong morphological signatures predicted by simulations of disc mergers 
with a range of mass ratios and orbital parameters.   

Translating the number of observed merger candidates into a merger rate requires the assumption
of an observability time-scale -- the time during which one would have identified the
system as merging.  Until now, the time-scales for detected disturbed morphologies have been poorly constrained. 
The observability time-scale for a particular merger may depend on (1) the method used to identify the
merger; (2) the merger parameters (mass ratio, gas properties, bulge/disc ratio, orbits, dust content);
and (3) the observational selection (observed wavelength, viewing angle, spatial resolution).
Cosmological-scale numerical simulations currently do not have the spatial resolution to directly determine the 
cosmologically-averaged observability time-scale for each method.  Therefore, one is required to
 use a suite of galaxy-scale numerical simulations which span a large range of input merger parameters to 
constrain the observability time-scales for the different input parameters. 
Given a sufficiently broad range of merger parameters,  the observability time-scales for each parameter set
may then be weighted by the probability distribution of the mass ratios, gas fraction, etc. which can be computed
from current cosmological-scale simulations.  An additional complication is that 
galaxy-scale numerical simulations typically 
track the distribution of `particles' (star, gas, and dark matter), as opposed to 
the projected light distribution at a particular wavelength (which is what is observed). 

We continue the work first presented in Lotz et al. 2008b,  hereafter Paper 1.   In Paper 1, we used {\sc GADGET} N-body/SPH simulations 
processed through the radiative transfer code {\sc SUNRISE} to obtain the rest-frame $g$ morphologies of equal-mass gas-rich disc mergers.
We found that the time-scales for identifying a particular galaxy merger could be quite different depending on the gas-fraction of the merging galaxies 
and the method used to find the merger.  For a given method, the time-scales associated with strong morphological disturbances are not a 
strong function of orbit, orientation, total mass, or supernovae feedback. The effects of numerical resolution, dust, viewing angle, and spatial 
resolution are also presented in Paper 1.  

In this paper,  we focus on the effect of mass ratio on the quantitative morphologies and projected separations of simulated mergers of disc galaxies
with gas fractions, bulge-to-disc ratios,  and masses tuned to match local galaxies.   
In \S 2, we describe the simulations, 
 the properties of the initial galaxies, and the range of merger parameters. In \S 3, we briefly
describe the analysis of the resulting simulated $g$-band images and the criteria for identification as a merger by morphology
and projected separation.  In \S 4, we discuss the resulting observability time-scales and their dependence on the merger mass ratios, orbits, and
relative orientation.  We will present the effects of gas fraction in a companion paper (Paper 3; Lotz et al. 2009). 
Those familiar with our approach from Paper 1 may skip to \S4$-$5 for the results and discussion.   The simulated $g$ band images and morphology
tables will be available in 2010 at the Multimission Archive at STScI (MAST) as a High-Level Science Products (HLSP) contribution ``Dusty Interacting Galaxy GADGET-SUNRISE
Simulations'' (DIGGSS):  {\bf http://archive.stsci.edu/prepds/diggss} . 

\begin{figure*}
\includegraphics[width=184mm]{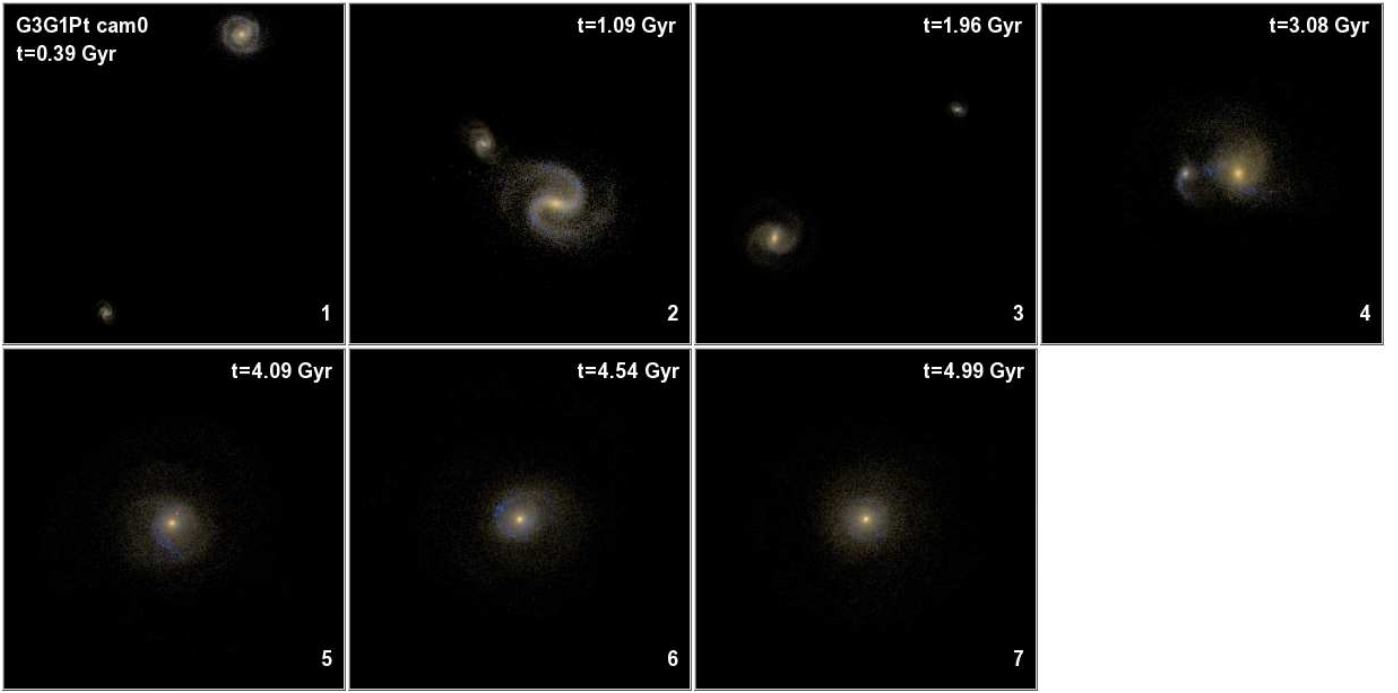}
\caption{$u-r-z$ composite colour images including dust extinction 
for the 9:1 baryonic mass ratio tilted prograde-prograde  G3G1P$_t$ simulation.   The viewing angles, merger stages and image
scales are the same as Figure \ref{g3g2col}.} \label{g3g1col}
\end{figure*}

\section{Simulations}
Here we briefly describe the galaxy merger simulations and initial conditions.   

\begin{table*}
  \centering
  \begin{minipage}{168mm}
    \caption{Initial Galaxy Conditions}
    \begin{tabular}{@{}lccrllllccccr@{}} 
      \hline
       Model  &  $N_{part}$\footnote{Number of particles in total, dark matter, gas, stellar disc, and stellar bulge for GADGET simulations} & $M_{vir}$\footnote{Virial mass} & $C$\footnote{Dark matter halo concentration}  & $M_{bary}$\footnote{Baryonic mass} & $M^*_{disc}$\footnote{Mass of stellar disc} & $M^*_{b}$\footnote{Mass of stellar bulge} & $M_{gas}$\footnote{Mass of gaseous disc} & $f_{b}$\footnote{Fraction of baryons in the bulge} & $f_{gas}$\footnote{Fraction of baryons in gas} & $R_{disc}$\footnote{Scalelength of stellar disc}  & $R_{b}$\footnote{Scalelength of bulge}  &$R_{gas}$\footnote{Scalelength of gaseous disc} \\
     &  & ($\Msun$) &  &($\Msun$) &($\Msun$) & ($\Msun$) & ($\Msun$)  &  &   & (kpc)  & (kpc)  & (kpc) \\
     \hline
 G3   & $2.4, [1.2, 0.5, 0.5, 0.2] \cdot10^5$  & $1.2\cdot10^{12}$  &6   & $6.2\cdot10^{10}$  &$4.1\cdot10^{10}$  &$8.9\cdot10^9$   &$1.2\cdot10^{10}$ & 0.14 &0.19  &2.85  & 0.62   &  8.55 \\
 G2   & $1.5, [0.8, 0.3, 0.3, 0.1] \cdot10^5$  & $5.1\cdot10^{11}$  &9   & $2.0\cdot10^{10}$  &$1.4\cdot10^{10}$  &$1.5\cdot10^9$   &$4.8\cdot10^{9}$  & 0.08 &0.24  &1.91  & 0.43   &  5.73 \\
 G1   & $9.5, [5.0, 2.0, 2.0, 0.5] \cdot10^4$  & $2.0\cdot10^{11}$  &12  & $7.0\cdot10^{9}$   &$4.7\cdot10^{9}$   &$3.0\cdot10^8$   &$2.0\cdot10^{9}$  & 0.04 &0.29  &1.48  & 0.33   &  4.44 \\
 G0   & $5.1, [3.0, 1.0, 1.0, 0.1] \cdot10^4$  & $5.1\cdot10^{10}$  &14  & $1.6\cdot10^{9}$   &$9.8\cdot10^{8}$   &$2.0\cdot10^7$   &$6.0\cdot10^{8}$  & 0.01 &0.38  &1.12  & 0.25   &  3.36 \\
\hline
\end{tabular}\label{gparstab}
\end{minipage} 
\end{table*}

\begin{figure*}
\includegraphics[width=184mm]{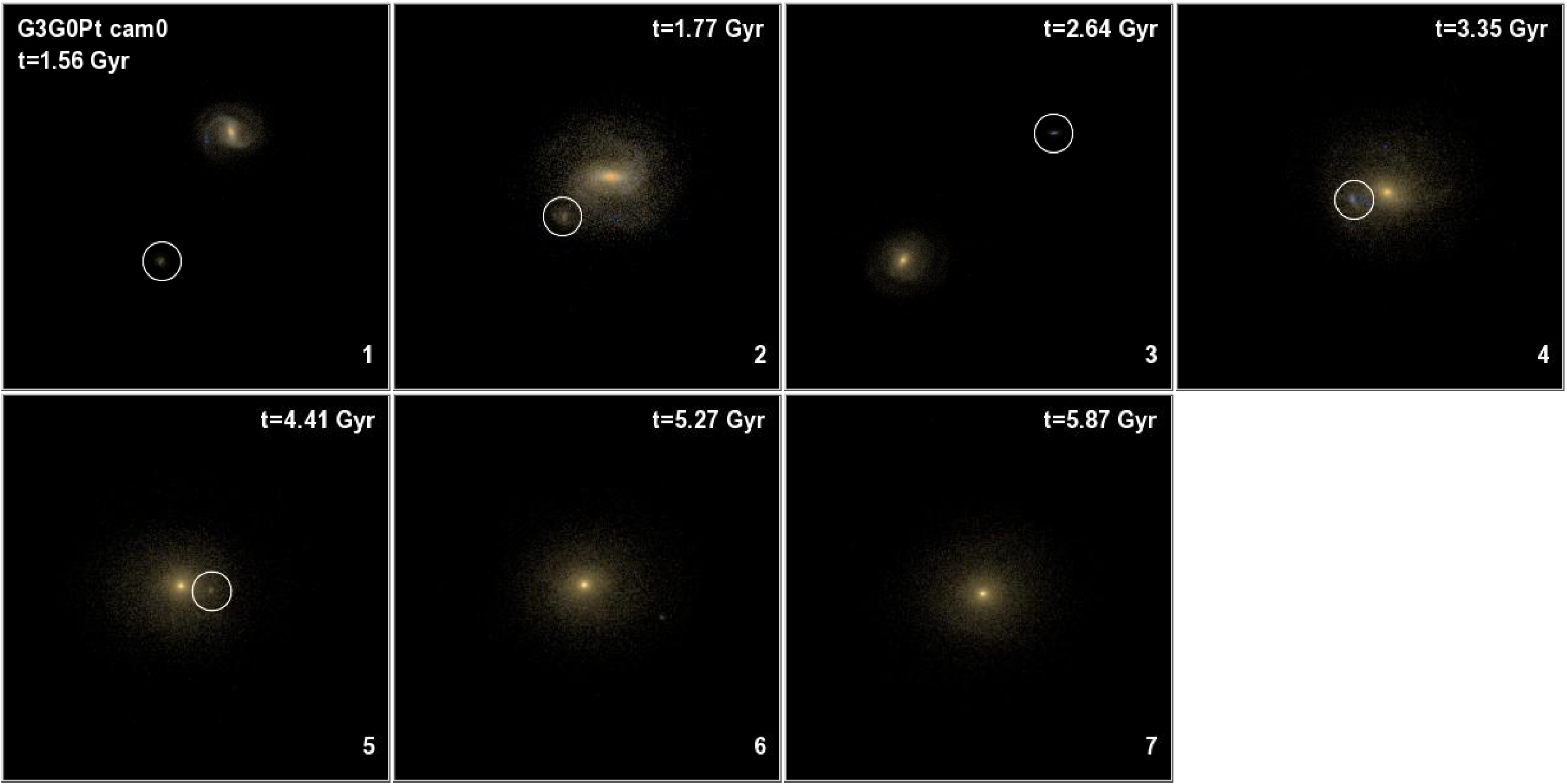}
\caption{$u-r-z$ composite colour images including dust extinction 
for the 39:1 baryonic mass ratio tilted prograde-prograde  G3G0P$_t$ simulation. The G0 secondary galaxy is stripped and fades considerably, 
and its position is circled.   G0 is almost completely disrupted by the fourth pass. 
The images show the initial galaxies [1], the first pass [2], the maximal separation after 
the first pass [3], the second pass [4],  the third pass [5], the fourth pass [6], and the last timestep calculated [7]. The viewing angles and image
scales are the same as Figure \ref{g3g2col}.} \label{g3g0col}
\end{figure*}

\subsection{{\sc GADGET} N-Body/SPH simulations}
 The details of these simulations, 
their global star-formation histories, and their remnant properties are discussed
in Cox et. al (2004, 2006, 2008). 
All of the simulations presented in this work were performed using the 
N-Body/SPH code {\sc GADGET} (Springel, Yoshida, \& White 2001). 
While we use the first version of GADGET (Springel et al. 2001), the smoothed
particle hydrodynamics are upgraded to use the `conservative entropy' version that
is described in Springel \& Hernquist (2002). Each galaxy is initially modeled as a disc of
stars and gas, a stellar bulge, and a dark matter halo, with the number of particles and masses
for each component given in Table 1.  The stellar and dark matter particles
are collisionless and are subject to only gravitational forces.  The gas particles are also
subject to hydro-dynamical forces. The baryonic and dark matter particles have gravitational
softening lengths of 100 pc and 400 pc respectively.  The SPH smoothing length for the
gas particles indicates the size of the region over which the particle's hydrodynamic quantities are
averaged and is required to be greater than half the gravitational softening length or
$>$ 50 pc.  The radiative cooling rate $\Lambda_{net}$($\rho$, $u$)
is computed for a primordial plasma as described in Katz et al. (1996).  

Gas particles are transformed into collisionless star particles assuming the
Kennicutt-Schmidt law (Kennicutt 1998) where the star-formation rate depends
on the local gas density $\rho_{gas}$. 
 These new star particles have typical masses $\sim 10^5$ M$_{\odot}$, and are assigned
ages based on their formation time and metallicities based on the metallicity of
the gas particle from which they are spawned.  We adopt the instantaneous recycling approximation
for metal production whereby massive stars are assumed to instantly become supernovae, 
and the metals produced are put back into the gas phase of the particle. 
In this version of GADGET, metals do not mix and remain in the gas particle in which they
are formed. The enriched gas contribution from stellar winds and Type Ia supernovae are ignored. 
Unlike the metals, there is no recycling of hydrogen and helium to the gas.  

Feedback from supernovae is required to produce stable star-forming discs. 
We adopt a model in which the supernova feedback energy is
dissipated on an 8 Myr time-scale, and have a equation of state  
$P \sim \rho_{gas}^{2}$.  No active galactic nuclei (AGN) are included in these simulations.  As we discussed in Paper 1, 
the exclusion of AGN feedback will not affect the morphological disturbance time-scales calculated here but may
affect the appearance of the merger remnants.

\subsection{{\sc SUNRISE} Monte Carlo radiative transfer processing}
{\sc SUNRISE} is a parallel code which performs full Monte Carlo radiative-transfer calculations
using an adaptive-mesh refinement grid (Jonsson 2006; Jonsson et al. 2006; Jonsson, Groves, \& Cox 2009).  
We use {\sc SUNRISE v2} to create $g$-band images at least 30 timesteps for each merger simulation.  
For each {\sc GADGET} simulation timestep,  {\sc SUNRISE} assigns a spectral energy distribution
to each star particle using the STARBURST99 population synthesis models (Leitherer et al. 1999).  

The metallicities of the gas and stars of the initial galaxy models decline
exponentially with the radius of the disc.  The density of dust is linearly proportional
to the density of metals in the gas.  The central metallicities and gradients scale with the
mass of the galaxy.   The details of the inital galaxy metallicities, gradients, and dust extinctions are
given in Rocha et al. (2008). 
 
Given a particular simulation geometry and viewing angle, {\sc SUNRISE v2}
performs the Monte-Carlo radiative transfer calculation for 20 wavelengths from the far-ultraviolet to 
the mid-infrared and interpolates a resulting spectral energy distribution of 510 wavelengths including the 
effects of absorption and scattering.  Images are created for eleven isotropically positioned viewpoints (``cameras'').
In Figures \ref{g3g2col}, \ref{g3g1col}, and \ref{g3g0col}, 
we show examples of composite $u-r-z$ images for the G3G2, G3G1, and G3G0 tilted prograde-prograde
simulation viewed face-on (camera 0)  at multiple timesteps. 

\subsection{Initial Galaxy Models}
In order to sample the parameter space spanned by many present-day galaxies, we explored mergers between model
galaxies with masses, bulge-to-disc ratios, and gas fractions motivated by SDSS estimates of typical local galaxies
 (Table \ref{gparstab}). Each galaxy model contains a rotationally supported disc of gas and stars, a non-rotating stellar bulge, and
a dark matter halo (Table \ref{gparstab}). A detailed description of the galaxy disc models can be found in Cox et al. (2006, 2008), 
Jonsson et al. (2006) and Rocha et al. (2008).

The largest galaxy (G3) is chosen to have a stellar mass $5 \times 10^{10}$ M$_{\odot}$, and the smaller galaxies
are chosen to have stellar masses $1.5 \times 10^{10}$ M$_{\odot}$ (G2),  $5 \times 10^{9}$ M$_{\odot}$ (G1),
and $1 \times 10^{9}$ M$_{\odot}$ (G0), spanning a factor of 50 in stellar mass.  
The stellar half-light radii are from the stellar mass-size relation
of Shen et al. (2003).  The bulge-to-disc ratios are taken from de Jong (1996) and used to determine the
stellar disc and bulge masses and scalelengths.   The gas fractions and masses are determined from the gas mass
- stellar mass scaling relation from Bell et al. (2003).  The gas scalelength is
assumed to be three times the stellar disc scalelength (Broeils \& Woerden 1994).   We adopt NFW dark matter halo profiles selected
such that the rotation curves lie on the baryonic Tully-Fisher relation (Bell \& de Jong 2001; Bell et al. 2003).
These models do not include adiabatic contraction. The total mass-to-light ratio is
assumed to vary with mass such that lower mass galaxy model have higher mass-to-light ratios.  

\begin{table}
  \centering
  \begin{minipage}{84mm}
    \caption{Merger Mass Ratios}
    \begin{tabular}{@{}ccccc@{}} 
      \hline
      Primary  &  Satellite   & Total   & Stellar  & Baryonic \\
     \hline
      G3       &    G3        & 1:1     & 1:1      & 1:1\\
      G3       &    G2        & 2.3:1   & 3.3:1    & 3.1:1\\
      G3       &    G1        & 5.8:1   & 10.0:1   & 8.9:1\\
      G3       &    G0        & 22.7:1  & 50.0:1   & 38.9:1 \\
     \hline
      G2       &    G2        & 1:1     & 1:1      & 1:1\\
      G2       &    G1        & 2.6:1   & 3.0:1    & 2.8:1 \\
      G2       &    G0        & 10.0:1  & 15.0:1   & 12.4:1 \\
     \hline
      G1       &    G1        & 1:1     & 1:1      & 1:1  \\
      G1       &    G0        & 3.9:1   & 5.0:1    & 4.4:1 \\
     \hline
\end{tabular}\label{mrtab}
\end{minipage}
\end{table} 

\begin{table}
  \centering
  \begin{minipage}{84mm}
  \caption{ Merger Simulation Parameters}
  \begin{tabular}{@{}lrrrrrr@{}}
    \hline
    Simulation & $\theta_1$\footnote{Initial orientation of galaxy 1 with respect to the plane of the orbit in spherical coordinates, 
    where $\theta = \arctan (\frac{\sqrt{ x^2 + y^2}}{z})$ and $\phi = \arctan (\frac{y}{x}$)} & $\phi_1^c$ 
    & $\theta_2$\footnote{Initial orientation of galaxy 2} & $\phi_2^d$   
    & $e$\footnote{Eccentricity of the orbit, where a parabolic orbit has $e=1$.} & R$_{peri}$\footnote{Pericentric distance of the initial orbit} \\
                    & (deg)  & (deg)  &  (deg)  & (deg) &    & (kpc) \\
     \hline
     \multicolumn{7}{c}{Equal-mass prograde-prograde mergers} \\
     \hline    
     G3G3P$_t$\footnote{In Paper 1, the equal-mass tilted prograde-prograde mergers are named G3PP, G2PP, and G1PP }   &-30 & 0     &30 & 60  &0.95  &13.6  \\
     G2G2P$_t^e$            &-30 & 0     &30 & 60  &0.95  &3.8  \\
     G1G1P$_t^e$            &-30 & 0     &30 & 60  &0.95  &3.0  \\
    \hline
     \multicolumn{7}{c}{Unequal-mass prograde-prograde mergers} \\
     \hline
     G3G2P$_t$         & -30  & 0	& 30 	& 60  & 0.95  & 13.6   \\
     G3G1P$_t$         & -30  & 0	& 30 	& 60  & 0.95  & 13.6   \\
     G3G0P$_t $        & -30  & 0    & 30    & 60  & 0.95    & 13.6  \\
     G2G1P$_t$         & -30  & 0   & 30    & 60	& 0.95  & 3.82  \\
     G2G0P$_t $        & -30  & 0   & 30    & 60	& 0.95  & 3.82  \\
     G1G0P$_t$         & -30  & 0   & 30    & 60	& 0.95  & 2.96   \\
     \hline
    \multicolumn{7}{c}{3:1 baryonic mass ratio mergers}  \\
    \hline
     G3G2P             & 0    & 0	& 30 	& 60  & 0.95  & 13.6  \\
     G3G2P$_-$         &-30   & 0	& 30 	& 60  & 0.95  & 6.8   \\
     G3G2P$_+$         & -30  & 0	& 30 	& 60  & 0.95  & 27.2   \\
     G3G2Pl$_t$        & -60  & 0	& 30 	& 60  & 0.95  & 13.6	\\
     G3G2Pl           & -90  & 0	& 30 	& 60  & 0.95  & 13.6   \\
     G3G2RP            & 180  & 0	& 30 	& 60  & 0.95  & 13.6	   \\
     G3G2PR            & -30  & 0  & 150   & 0  & 0.95     & 13.6    \\
     G3G2C             &-30   & 0	& 30 	& 60  & 0.    & $-$ \\
     \hline
     \multicolumn{7}{c}{9:1 baryonic mass ratio mergers} \\
     \hline
     G3G1P             & 0	& 0	& 30 	& 60  & 0.95  & 13.6  \\
     G3G1P$_-$         & -30 & 0	& 30 	& 60  & 0.95  & 6.8   \\
     G3G1P$_+$         & -30 & 0	& 30 	& 60  & 0.95  & 27.2   \\
     G3G1Pl$_t$        & -60 & 0	& 30 	& 60  & 0.95  & 13.6	\\
     G3G1Pl            & -90 & 0	& 30 	& 60  & 0.95  & 13.6   \\
     G3G1RP            & 180 & 0	& 30 	& 60  & 0.95  & 13.6	   \\
     G3G1PR            & -30 & 0   & 150   & 0   & 0.95  & 13.6    \\
     G3G1C             & -30 & 0	& 30 	& 60  & 0.    & $-$ \\
\hline
   \end{tabular} \label{simpartab}
 \end{minipage}
\end{table}

\subsection{Merger Parameters}
The new galaxy merger simulations presented here are mergers of discs spanning a range of mass ratios and
merger orbits and orientations. Mergers of equal-mass gas-rich mergers were previously presented in Paper 1.
The total, baryonic, and stellar mass ratios are given in Table \ref{mrtab}.  We will generally
refer the baryonic mass ratios of the merger, but note that the stellar/total mass ratios
are higher/lower than the baryonic mass ratios.  These ratios range from 1:1 $-$ 23:1 for
total mass ratio,  1:1 $-$ 39:1 for baryonic mass ratio, and 1:1$-$50:1 for stellar mass ratio.  
The G3G3, G2G2, G1G1 mergers have 1:1 mass ratios.  The G3G2 and G2G1 mergers have baryonic mass ratios of $\sim$ 3:1  
and hence are more representative of the typical major merger than rare equal-mass mergers.   
The G3G1, G3G0, G2G0, and G1G0 mergers have baryonic mass ratios of 9:1, 39:1, 12:1, and 4:1 respectively, 
and are considered to be minor mergers. 

Each permutation of primary-satellite merger has been simulated with a tilted prograde-prograde orientation with an initial sub-parabolic
orbit (labeled P$_t$) with a pericentric radius $R_{peri} = 13.6$ kpc. Additional simulations of major G3G2 mergers and minor G3G1 
mergers have been performed with a wide range of orbital parameters (Table \ref{simpartab}).  These include tilted prograde-prograde orientations on sub-parabolic 
orbit with two additional pericentric distances (P$_-$: $R_{peri} = 6.8$ kpc, P$_+$: $R_{peri} = 27.2$ kpc),  
pure prograde-prograde orientation (P), pure and tilted polar orientations (Pl, Pl$_t$), 
primary retrograde $-$ satellite prograde orientation (RP), primary prograde $-$ satellite retrograde orientation (PR), and
circular orbit  (C;  see Table \ref{simpartab}).  All of simulations initially have slightly sub-parabolic orbits with an eccentricity $e = 0.95$, except
for the circular orbits simulations, which by definition have $e=0$.   All of the parabolic orbit simulations start with an initial separation of 250 kpc; 
the circular orbit simulations start with an initial separation of 120 kpc. 

For comparison, we have also calculated the morphologies for an isolated G3 galaxy.  The isolated G3 simulation was run for
6 Gyr.  Because no additional gas is provided, the isolated G3 galaxy consumes its initial gas reservoir and fades as the 
simulation progresses.  

\section{Image Analysis}
We replicate the observations and measurements of real galaxy mergers as closely 
as possible.   We focus on rest-frame $g$ morphologies for purposes of this paper, as
these simulations can be used to calibrate the morphologies of galaxies currently observed from the ground and
with the Hubble Space Telescope in optical and near-infrared wavelengths at $0 < z < 3$. 
In the following section we briefly describe how the simulated $g$ images are degraded and analysed to match real
galaxy morphology measurements; a more detailed discussion may be found in Paper1. 

\subsection{Image degradation}
The images are produced by {\sc SUNRISE} for each simulation for 11 isotropically positioned viewpoints as a function
of time from $\sim 0.5$ Gyr prior to the first pass to $\ge$ 1 Gyr after the final coalescence in $\sim 30-250$ 
Myr timesteps depending on the merger state, up to a maximum runtime of 6 Gyr.   Only the G3G1PR and G3G1C simulations 
have not coalesced by this time.  The nuclei of the G3G0P$_t$ simulation don't truely merge,  but the G0 satellite is almost
completely disrupted by its fourth pass.  The field of view of the output images ranges from 200 kpc 
during the initial stages and period of  maximal separation to 100 kpc during the first pass, second pass, 
final merger and post-merger stages.  The intrinsic resolution of the output SUNRISE $g$-band images is 333 pc per pixel.

The images output by {\sc SUNRISE} have no background sky noise and no seeing effects, although they
do have particle noise and Monte Carlo Poisson noise. We degrade these images to simulate real data, 
but do not attempt to mimic a particular set of galaxy survey observations.  
We re-bin the images to 105 pc per pixel and convolve the images with a Gaussian function with a FWHM = 400 pc.  
This was done to simulate the effect of seeing but maintain as high spatial resolution as possible.  The values where
chosen to match the Sloan Digital Sky Survey  (Abazajian et al. 2003) with 1.5\arcsec\ seeing, 0.396\arcsec\ per pixel plate scale for a galaxy at a distance such
that 1.5\arcsec\ $\sim$ 400 pc.  We also add random Poisson noise background to simulate sky noise but scale this noise
to maintain a high signal-to-noise for the primary galaxies ($>20$ per pixel within the Petrosian radius).   For the G3G0 simulation, 
the G0 satellite galaxy fades substantially because of tidal stripping and declining star-formation,  and has average signal-to-noise per pixel $< 3$
after the third pass. 

 \begin{table}
  \centering
  \begin{minipage}{84mm}
  \caption{Morphological Disturbance Time-Scales}
  \begin{tabular}{@{}lccc@{}}
 \hline
  Simulation &  T($G-M_{20}$)  & T($G-A$)  & T($A$) \\
                    &   (Gyr)                &   (Gyr)          & (Gyr)  \\ 
                    
\hline
\multicolumn{4}{c}{Equal-mass prograde-prograde mergers} \\
\hline    
G3G3P$_t$ &  0.16$\pm$ 0.07 &  0.31$\pm$ 0.08 &  0.23$\pm$ 0.11   \\
G2G2P$_t$ &  0.22$\pm$ 0.14 &  0.31$\pm$ 0.15 &  0.25$\pm$ 0.19   \\
G1G1P$_t$&  0.23$\pm$ 0.04 &  0.35$\pm$ 0.14 &  0.30$\pm$ 0.16   \\
\hline
\multicolumn{4}{c}{Unequal-mass prograde-prograde mergers} \\
\hline    
G3G2P$_t$ &  0.25$\pm$ 0.08 &  0.30$\pm$ 0.13 &  0.24$\pm$ 0.11   \\
G3G1P$_t$ &  0.36$\pm$ 0.15 &  0.27$\pm$ 0.13 &  0.03$\pm$ 0.03   \\
G3G0P$_t$ &  0.03$\pm$ 0.03 &  0.02$\pm$ 0.03 &  0.00$\pm$ 0.00   \\
G2G1P$_t$ &  0.17$\pm$ 0.06 &  0.30$\pm$ 0.09 &  0.26$\pm$ 0.10   \\
G2G0P$_t$ &  0.10$\pm$ 0.08 &  0.06$\pm$ 0.05 &  0.01$\pm$ 0.02   \\
G1G0P$_t$ &  0.11$\pm$ 0.06 &  0.22$\pm$ 0.10 &  0.14$\pm$ 0.07   \\
\hline
\multicolumn{4}{c}{3:1 baryonic mass ratio mergers} \\
\hline    
G3G2P    &  0.20$\pm$ 0.10 &  0.28$\pm$ 0.06 &  0.25$\pm$ 0.06   \\
G3G2P$_-$   &  0.32$\pm$ 0.10 &  0.39$\pm$ 0.12 &  0.30$\pm$ 0.12   \\
G3G2P$_+$   &  0.20$\pm$ 0.14 &  0.27$\pm$ 0.10 &  0.18$\pm$ 0.11   \\
G3G2Pl$_t$ &  0.30$\pm$ 0.14 &  0.36$\pm$ 0.09 &  0.25$\pm$ 0.09   \\
G3G2Pl   &  0.34$\pm$ 0.11 &  0.47$\pm$ 0.14 &  0.37$\pm$ 0.10   \\
G3G2RP   &  0.38$\pm$ 0.14 &  0.43$\pm$ 0.11 &  0.35$\pm$ 0.11   \\
G3G2PR   &  0.28$\pm$ 0.11 &  0.44$\pm$ 0.17 &  0.29$\pm$ 0.07   \\
G3G2C    &  0.31$\pm$ 0.17 &  0.40$\pm$ 0.20 &  0.36$\pm$ 0.17   \\
\hline
\multicolumn{4}{c}{9:1 baryonic mass ratio mergers} \\
\hline
G3G1P      &  0.24$\pm$ 0.11 &  0.21$\pm$ 0.13 &  0.03$\pm$ 0.03   \\
G3G1P$_-$  &  0.26$\pm$ 0.07 &  0.26$\pm$ 0.14 &  0.06$\pm$ 0.04   \\
G3G1P$_+$  &  0.46$\pm$ 0.26 &  0.24$\pm$ 0.06 &  0.02$\pm$ 0.03   \\
G3G1Pl$_t$ &  0.37$\pm$ 0.13 &  0.26$\pm$ 0.15 &  0.01$\pm$ 0.04   \\
G3G1Pl     &  0.32$\pm$ 0.08 &  0.28$\pm$ 0.14 &  0.06$\pm$ 0.05   \\
G3G1RP     &  0.22$\pm$ 0.07 &  0.11$\pm$ 0.10 &  0.06$\pm$ 0.05   \\
G3G1PR\footnote{The simulation does not merge within 6 Gyr.}   & $\ge$  0.26$\pm$ 0.13 & $\ge$  0.08$\pm$ 0.08 & $\ge$  0.03$\pm$ 0.07   \\
G3G1C$^a$     & $\ge$   0.05$\pm$ 0.06 &  $\ge$  0.00$\pm$ 0.01 &  $\ge$  0.00$\pm$ 0.00   \\
\hline
\end{tabular}\label{tmorphtab}
\end{minipage}
\end{table}  

\subsection{Morphology Measurements}
Each image is run through an automated galaxy detection algorithm integrated into our 
IDL code. If the centres of the merging galaxies are less than 10 kpc apart, they are generally detected as a single object. 
If 2 distinct galaxies are detected, the detection segmentation maps are used to mask out the other galaxy while each 
galaxy’s morphology is measured.   The projected separation $R_{proj}$ is measured when two galaxies are detected.
For each detected object, we calculate the Petrosian radii in circular and elliptical apertures, concentration $C$ , 
180 degree rotational asymmetry $A$,  the Gini coefficient $G$, and the second-order moment of the brightest 20\% of the 
light $M_{20}$. (Please refer to Lotz, Primack, \& Madau 2004, Conselice 2003, and Paper 1 for detailed definitions).  

In Paper 1, we studied the effect of numerical resolution on the simulation morphologies.  We found small but significant differences in 
the average $M_{20}$ and $A$ values when we compared the standard numerical resolution simulation to a simulation with 10 times as many
particles, and corrected for these offsets. Because the numerical resolution of the simulations presented here are similar to the 
standard resolution simulations in Paper 1  ($\sim 10^5$ particles), we apply the same correction of $\delta M_{20} = -0.157$ and $\delta A = -0.115$ to 
the values in this work.  

\begin{figure*}
\includegraphics[width=168mm]{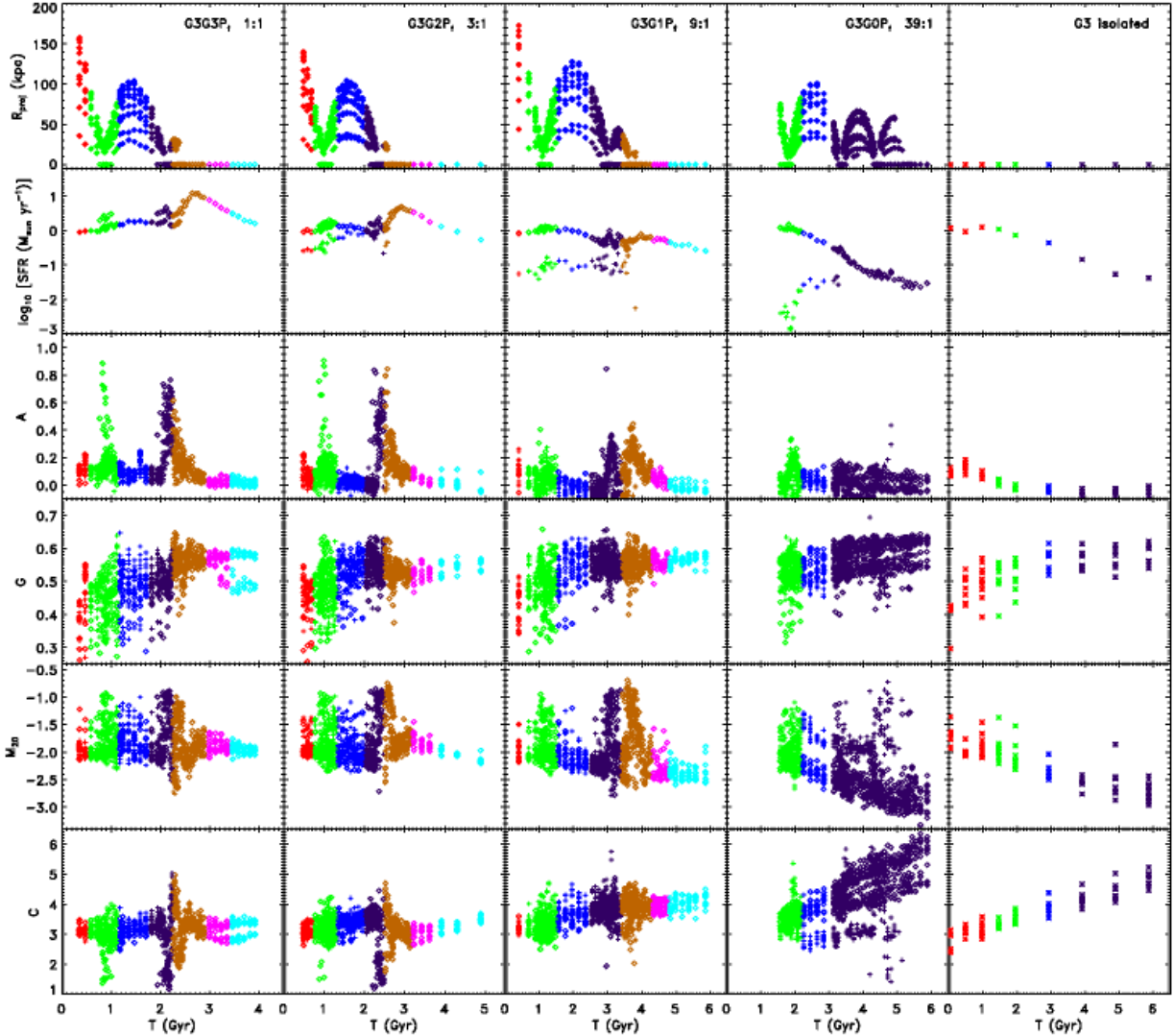}
\caption{Time v. projected separation ($R_{proj}$), log$_{10}$(SFR), and morphology ($A$, $G$, $M_{20}$, $C$) 
for G3G3P$_t$,  G3G2P$_t$, G3G1P$_t$, G3G0P$_t$ mergers and isolated G3 galaxy. The satellite galaxies are 
shown as crosses, and the primary galaxy/merger is shown as diamonds. The merger stages are colour coded
with initial galaxies:red, the first pass:green, the maximal separation:blue, second pass (and third pass): purple, final merger:orange, 
post-merger:magenta, and merger remnant at $>$ 1 Gyr after coalescence of the
nuclei as cyan.  The isolated G3 points are colour-coded to match the G3G0P$_t$ simulation. 
The simulations show peaks in star-formation and morphological disturbances at the first pass
and final merger,  with short-lived or no asymmetries for the minor mergers G3G1 and G3G0. }  \label{tmrg3}
\end{figure*}

\begin{figure*}
\includegraphics[width=168mm]{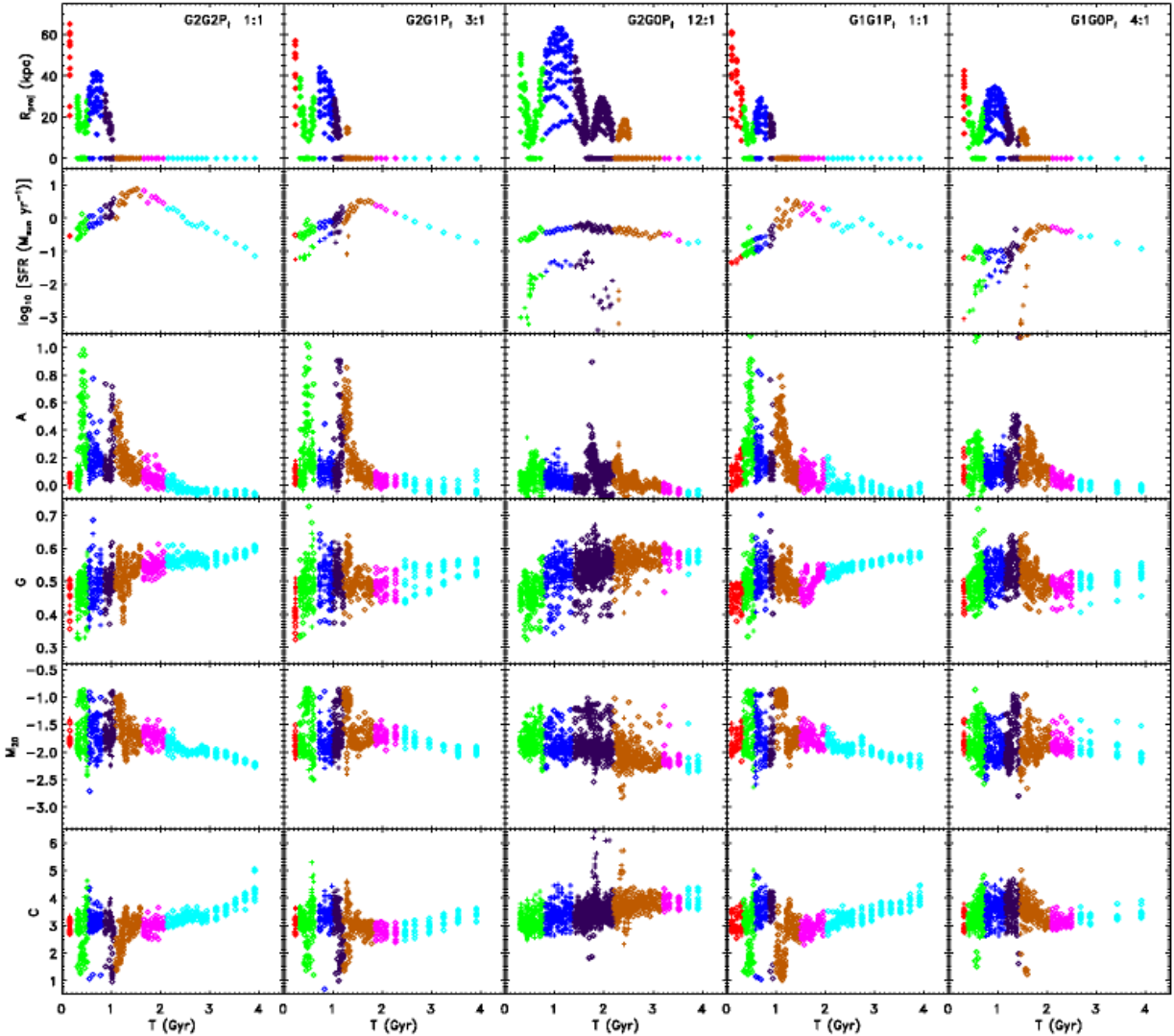}
\caption{Time v. projected separation ($R_{proj}$), log$_{10}$(SFR), and morphology ($A$, $G$, $M_{20}$, $C$) 
for lower-mass G2G2P$_{t}$,  G2G1P$_{t}$, G2G0P$_{t}$, G1G1P$_t$ and G1G0P$_{t}$ mergers. The satellite galaxies are 
shown as crosses, and the primary galaxy/merger is shown as diamonds. The merger stages 
are colour coded as in the previous figure. As for the G3G1P$_t$ minor merger, the minor mergers G2G0P$_t$ and G1G0P$_t$ 
appear asymmetric for a much shorter period of time than the major mergers G2G2P$_{t}$, G2G1P$_t$, and G1G1P$_t$. }  \label{tmrg21}
\end{figure*} 

\subsection{Merger Classification and Time-Scales}
Lotz et al. (2004) found that local ultra-luminous galaxies visually classified as mergers could be distinguished from
the sequence of normal Hubble type galaxies with  
\begin{equation}
G > -0.115\ M_{20} + 0.384
\end{equation}
or
\begin{equation}
G  > -0.4\ A + 0.66\ {\rm or}\  A \ge 0.4
\end{equation}

Asymmetry alone is also often used to classify merger candidates.   The calibration of local mergers by
Conselice (2003) finds the following merger criterion:
\begin{equation}
A \ge 0.35
\end{equation} 

Galaxies at higher redshift cannot be imaged at as high spatial resolution as local galaxies even
when observed with $HST$.  The measured morphologies of galaxies at $z > 0.25$ imaged with $HST$ will have
non-negligible biases as a result of this lower spatial resolution (Lotz et al. 04).  See Paper 1 for a
discussion of how the morphologies and merger criteria change for $z \sim 1$ $HST$ resolution, and
how the time-scale computed here may be applied to $HST$ data.  

Close kinematic pairs are also probable pre-merger systems.  
We assume $h=0.7$ and we estimate the time-scales during which merging galaxies can be found as separate objects
within $5  < R_{proj} < $ 20, 10 $ < R_{proj} <$ 30, 10 $ < R_{proj} < $ 50, 
and 10 $< R_{proj} < $ 100 $h^{-1}$ kpc.    The simulated merging galaxies always have relative velocities $<$ 500  
km s$^{-1}$. 

We calculate each simulation's average observability time-scale for the $G-M_{20}$, $G-A$, and $A$ criteria given
above by averaging the results of the 11 isotropic viewpoints.  
Because we wish to determine the number density of merger events rather than the number of galaxies
undergoing a merger, galaxies that have not yet merged but identified morphologically as merger candidates are 
weighted accordingly.  The time that each pre-merger galaxy is morphologically disturbed is added (not averaged)
to the time that the post-merger system appears disturbed. 
No such weighting is done for the close pair time-scales as this factor is generally
included in the merger rate calculation (e.g. Patton et al. 2000).

\begin{figure*}
\includegraphics[width=168mm]{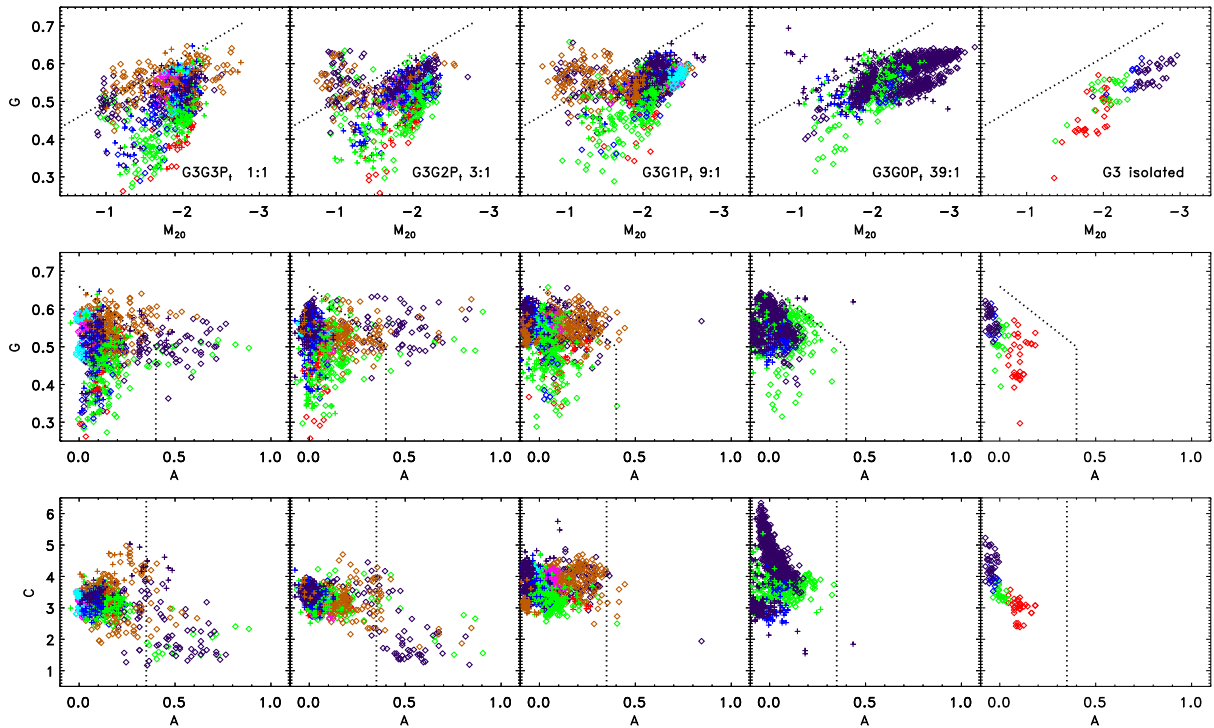}
\caption{$G-M_{20}$, $G-A$, and $C-A$ for  for G3G3P$_t$,  G3G2P$_t$, G3G1P$_t$, G3G0P$_t$ mergers and isolated G3 galaxy.
The satellite galaxies are shown as crosses, and the primary galaxy/merger is shown as diamonds.
The merger stages are colour-coded as in the previous plots.  The isolated G3 galaxy is colour-coded to match the stages of the G3G0 simulation.
The simulations show peaks in morphological disturbances at the first pass (green points),
second pass (purple points), and final merger (orange points),  with short-lived or no asymmetries for the minor mergers G3G1P$_t$ and G3G0P$_t$. } \label{morphg3}
\end{figure*}

\begin{figure*}
\includegraphics[width=168mm]{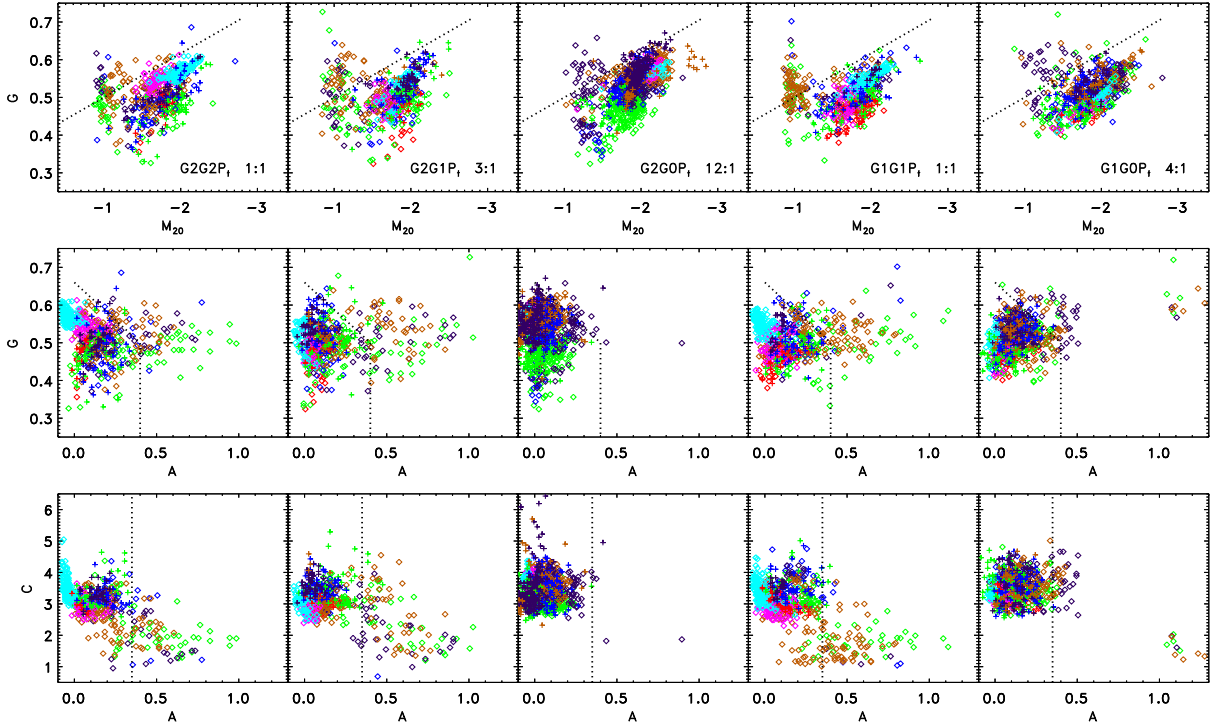}
\caption{$G-M_{20}$, $G-A$, and $C-A$ for lower-mass G2G2P$_t$,  G2G1P$_{t}$, G2G0P$_{t}$, G1G1P$_t$ and G1G0P$_{t}$ mergers. 
The satellite galaxies are shown as crosses, and the primary galaxy/merger is shown as diamonds. The merger stages are colour coded as in the
previous figure. As for the G3G1P$_t$ minor merger, the minor mergers G2G0P$_t$ and G1G0P$_t$  are asymmetric for a much shorter period of 
time than the major mergers G2G2P$_t$, G2G1P$_t$, and G1G1P$_t$. } \label{morphg21}
\end{figure*}

\begin{figure*}
\includegraphics[width=168mm]{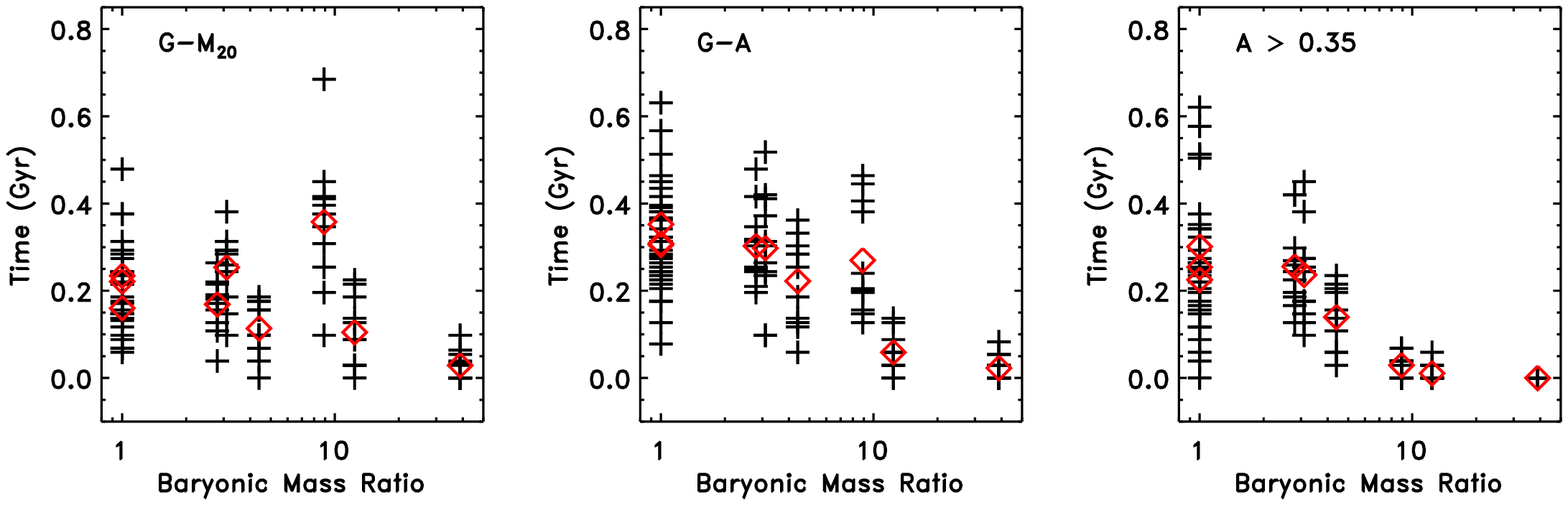}
\caption{Time-scales for morphological disturbances in $G-M_{20}$, $G-A$, and $A$ v. baryonic mass ratio. 
The black crosses show the time-scales for each viewing angle, and  the red diamonds are the viewing-angle averaged time-scales.
The time-scales for observing high asymmetry is a strong function of mass ratio, while the time-scales for observing 
objects with high $G-M_{20}$ values appears to be independent of mass ratio below baryonic mass ratio $\sim$ 10.  
Therefore $G-M_{20}$, unlike $A$, will identify minor mergers with baryonic mass ratios $\le$ 9:1 as well as major mergers.} \label{tmorphmr}
\end{figure*}

\section{Results}
\subsection{Evolution with Merger Stage}
The time evolution of each simulation depends on the mass ratios, masses, and orbits of the galaxies. Therefore it is useful
to compare the behavior of each simulation as a function of merger stage rather than time. 
We define seven different merger stages based on the positions of the galaxy nuclei in physical space, 
which are colour-coded in Figures \ref{tmrg3}$-$ \ref{tmrg21}. These are the initial encounter (red), 
the first pass (green),  maximal separation immediately after the first pass (blue), the second pass or final approach (purple),
the final merger (orange), the post-merger observed 0.5$-$1 Gyr after the final merger (magenta), and the remnant
observed $>$ 1 Gyr after the final merger (cyan). 

In Figures \ref{tmrg3} and \ref{tmrg21} we examine the projected separations $R_{proj}$, measured morphologies ($G$, $M_{20}$, $C$, and $A$), 
and star-formation rate per object as a function of time and merger stage for the tilted prograde-prograde simulations, spanning a range
of baryonic mass ratio of 1:1 $-$ 39:1 for the G3 simulations (Figure \ref{tmrg3}) and 1:1 $-$ 12:1 for the G2 and G1 simulations (Figure \ref{tmrg21}). 
 The scatter in $R_{proj}$ and morphologies at each timestep is the variation
in the merger appearance with the 11 viewing angles. The minor mergers G3G1P$_t$ and G2G0P$_t$ take significantly longer 
to complete than the major mergers, and experience a third and final pass before the galaxies coalesce.  
In the G3G0P$_t$, the G0 satellite is disrupted and becomes too faint to be
detected above the noise in our simulated images after the third pass so the projected separation falls to zero (Figure \ref{g3g0col}).  

The simulations generally show similar correlations of star-formation rate  and morphology with merger-stage.  
The initial segmentation maps computed to identify each galaxy are used to compute the total star-formation rate for each galaxy
at each timestep and camera. 
The major mergers experience a burst of star-formation at the first pass and a few hundred Myr after the galaxies merge as their
nuclei coalesce.  As we found in Paper1, this peak in star-formation rate at the final merger occurs a few hundred Myr after 
the strongest morphological disturbances in $A$, $C$, and $M_{20}$.

The minor mergers also have enhanced star-formation rates at the close passages.  However,  the minor mergers' burst efficiencies, or
stars formed in excess of what is expected for isolated galaxies, is lower than for major mergers (Cox et al. 2008;
see also Johansson et al. 2009).  The G3G1P$_t$ and G2G0P$_t$ mergers do not show strong starbursts at the final merger. 
This is because the systems have already consumed most of their gas by the time of the final merger $\sim 2-4$ Gyr after the first encounter. 
The G3G0P$_t$ simulation shows the strongest decline in the star-formation rate of all the G3 merger simulations. 
The G3G0 has the least gas available for star-formation because of the low gas mass of the G0 satellite,  and the evolution of
its star-formation rate with time is very similar to the isolated G3 galaxy (Figure \ref{tmrg3}). 
 
Except for the G3G0P$_t$ simulation,  all of the mergers show peaks in $A$ and 
$M_{20}$ and minima in $C$ during the close passages and just before the final merger, as was also found for the equal-mass mergers in Paper 1.  
However, mass ratio affects the strength of these morphological disturbances.   In Figure \ref{tmrg3}, we find that the major mergers G3G3P$_t$ and
G3G2P$_t$ have roughly similar offsets in $A$ and $C$ at the first pass and final approach/merger.   The minor merger G3G1P$_t$ has much weaker
disturbances in $A$ and $C$ at the first pass and final approach/merger,  while G3G0P$_t$ shows only a small increase in $A$ at the first pass. 
The increase in $M_{20}$, on the other hand, is equally strong for the G3G3P$_t$, G3G2P$_t$, and G3G1P$_t$ mergers.  In Figure \ref{tmrg21}, we see that the
lower mass G2 and G1 mergers show similar dependence on mass ratio.  The major mergers G2G2P$_t$, G2G1P$_t$, and G1G1P$_t$ have similarly strong peaks in
$A$ and minima in $C$ at the first pass and final approach/merger,  while the minor mergers G2G0P$_t$ and G1G0P$_t$ have weaker disturbances in $A$ and $C$. 

Dust extinction is a function of the metallicity and gas-fraction of the merging system, 
and has the strongest effect on $G$ (Paper 1).   
For the more massive G3 and G2 mergers with higher gas metallicity, $G$ shows significant scatter with
viewing angle between the first pass and final merger as a result of dust extinction along certain lines of sight.
The lowest mass G1 mergers show higher $G$ values between the first pass and final merger because this simulation has the lowest
metallicity and dust content.  

As in Paper 1,  we find that the major merger remnants observed $>$ 1 Gyr after the merger have $G$, $M_{20}$, $C$, $A$, and star-formation rates
more similar to local early-type spiral galaxies ($G \sim 0.55$, $M_{20} \sim -2.0$, $C \sim 3.5$, $A \sim 0.05$) than 
local elliptical/S0 galaxies ($G \sim 0.60$, $M_{20} \sim -2.5$, $C \sim 4.5$, $ A \sim 0.0$).  These merger remnants retain significant amounts
of gas and reform small-mass star-forming discs.   Despite their massive bulges,  these low-mass disc components are bright enough in $g$-band
to affect the remnants' quantitative morphologies (see also Barnes 2002, Springel et al. 2005, Naab et al. 2006, Robertson et al. 2006).    
We note that our simulations do not include any form of AGN feedback, which could remove this
remaining gas and prevent the formation of discs large enough to affect the remnant morphologies (e.g. di Matteo et al. 2005), although
AGN feedback may not be sufficient to quench star-formation in unequal-mass mergers (Johansson et al. 2009).   The supernovae feedback efficiency
and merger geometries can also effect the amount and spatial distribution of gas in the merger remnant (e.g. Springel et al. 2005, Cox et al. 2006,  Paper 1).

The minor merger remnants, on the other hand, do have quantitative morphologies similar to ellipticals/S0 galaxies.  
But these simulations have been run for several Gyr longer than the
major merger simulations without any additional gas supplied.   Visual inspection of the images reveals that the minor merger remnants have compact nuclei
and red discs rather than massive bulge components (Figures \ref{g3g1col}, \ref{g3g0col}).  Thus the difference in the morphologies for major and minor 
merger remnants is likely the result of
longer simulation runtimes/greater gas consumption for the minor merger simulations.   We find that an isolated G3 galaxy simulation run for 6 Gyr with
no additional gas supply results in an object very similar in appearance and quantitative morphology to the minor merger remnants (Figure \ref{tmrg3}).  Therefore we
conclude that minor gas-rich mergers have a minimal effect on the morphologies of their merger remnants.   The optical properties of the merger
remnants will be explored in more detail in a future paper.

\subsection{Morphology Time-Scales}
We calculate the viewing-angle averaged time each merger simulation exhibits morphologies which meet the 
merger criteria for $G-M_{20}$, $G-A$, and $A$ (Table \ref{tmorphtab}).    We find that the mass ratio
is an important factor in determining how long or whether a merger may be identified as a merger in $G-M_{20}$, 
$G-A$, or $A$, but the merging galaxy orientations and orbits are not.

The merger simulations clearly show different behaviors in $G-M_{20}$, $G-A$, and $A$ as a function of mass ratio. 
We plot $G-M_{20}$, $G-A$, and $C-A$ for the G3G3P$_t$, G3G2P$_t$, G3G1P$_t$, G3G0P$_t$ mergers (Figure \ref{morphg3}) and
G2G2P$_t$, G2G1P$_t$, G2G0P$_t$, G1G1P$_t$, and G1G0P$_t$ mergers (Figure \ref{morphg21}).   We find that a significant number of timesteps lie in the merger region 
of the $G-M_{20}$ plots for all the merger simulations except for the highest mass ratio G3G0P$_t$ simulation.  On the other hand, only the major mergers 
(G3G3P$_t$, G3G2P$_t$, G2G2P$_t$, G2G1P$_t$, and G1G1P$_t$) have a large number of timesteps in the merger regions of the $G-A$ and $A$ plots.  The
isolated G3 simulation never shows morphologies that meet the merger criteria (right panels, Figure \ref{morphg3}). 

In Figure \ref{tmorphmr}, we plot the observability time-scales v. baryonic mass ratio for the tilted prograde-prograde mergers.  
We find that $G-M_{20}$ shows no significant decrease in its observability time-scale for baryonic mass ratios $\leq$ 9:1, 
and is as likely to find 9:1 minor mergers as 1:1 major mergers. The asymmetry time-scales are stronger function of mass ratio,  
and $A > 0.35$ does not detect mergers with baryonic mass ratios $>$ 5:1.  

\begin{figure}
\includegraphics[width=84mm]{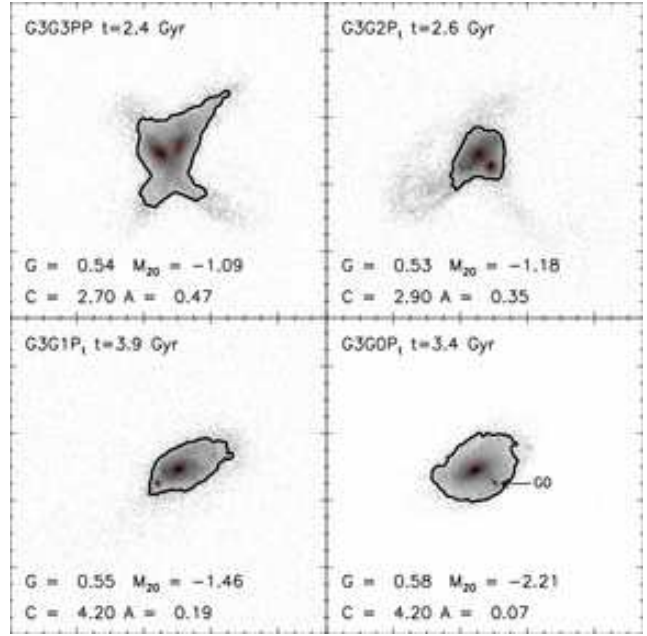}
\caption{The brightest 20\% of light is marked in red contours for G3G3P$_t$, G3G2P$_t$, G3G1P$_t$, and G3G0P$_t$ mergers just before the final coalesence
(except for G3G0P$_t$, which is shown at the second pass and does not merge).   The satellite nuclei are among the brightest regions for mergers with 
baryonic mass ratio $\leq$ 9:1 (G3G1),  but are too faint to be detected by $G-M_{20}$ at baryonic mass ratio 39:1 (i.e. G3G0). } \label{m20}
\end{figure}

We visually inspected the merger images to better understand why $G-M_{20}$ and $A$ show these different behaviours.  
In Figure \ref{m20},  we show the $g$-band images of tilted prograde-prograde mergers viewed 
just before the final merger (or at the second pass for G3G0P$_t$).  $M_{20}$ is a measure of the second-order moment of the
brightest 20\% of an object's flux, normalized by the total second-order moment. The brightest 20\% of the light is shown in Figure \ref{m20} 
by the inner red contours, and the outer black contour shows the segmentation map used to compute $G$ and $M_{20}$. 
The nucleus of the merging satellite is among the brightest 20\% of the light at this merger stage for all the mergers except for the highest 
mass ratio merger G3G0P$_t$.   Therefore, $G-M_{20}$ will detect mergers when the nucleus of the merging satellite is among the brightest 20\% of
the system's pixels. Asymmetry, on the other hand,  is a measure of the rotational asymmetry of the system, and depends on the sum  of the absolute
difference between the original image and the image rotated 180 degrees about the center of minimum asymmetry.  Therefore $A$ depends more
strongly on the relative brightnesses of the nuclei, and is largest when the nuclei are of similar luminosity. 

\begin{figure*}
\includegraphics[width=168mm]{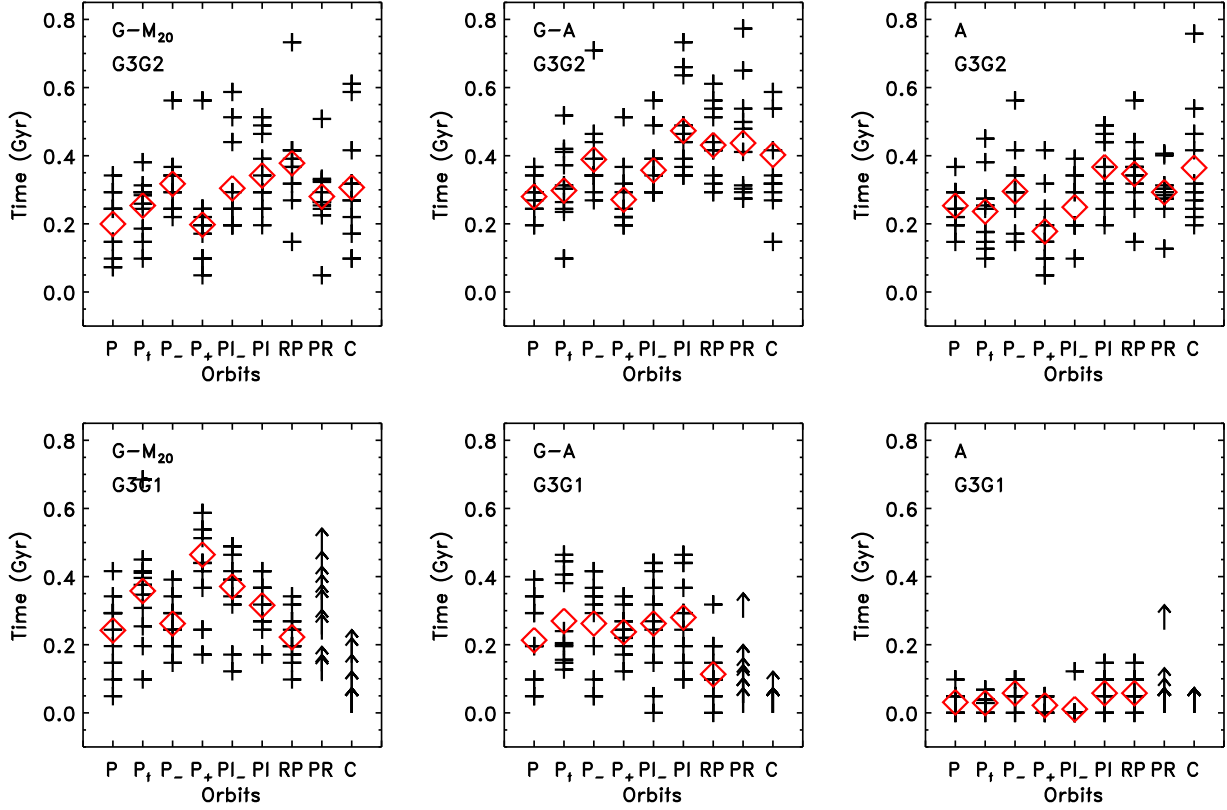}
\caption{Time-scales for morphological disturbances in $G-M_{20}$, $G-A$, and $A$ v. orientation and orbits 
for the 3:1 baryonic mass ratio G3G2 and 9:1 baryonic mass ratio 
G3G1 mergers (see Table \ref{simpartab}).  The time-scales do not depend strongly on the merger orientations or orbits.  
The black crosses show the time-scale for each viewing angle, and  the red diamonds are the viewing-angle averaged time-scales.
The G3G1PR and G3G1C simulations do not merge within 6 Gyr, hence the time-scales shown are lower limits (arrows).} \label{orbit}
\end{figure*}

As we found for equal-mass mergers in Paper 1, the orbits and relative orientations do not have a significant effect on the time-scales for 
identifying unequal-mass galaxy mergers for any of the morphology criteria.  In Figure \ref{orbit}, we plot the observability time-scales 
for $G-M_{20}$, $G-A$, and $A$ for the G3G2 and G3G1 mergers with nine different orbits and orientations. The G3G1 circular orbit (C)  
and G3 prograde $-$ G1 retrograde parabolic orbit (PR) simulations do not merge within simulation runtimes of 6 Gyr, and hence the plotted
time-scales are upper limits (arrows).  The mean observability time-scales 
for each orbit/orientation  (red diamonds) have less variation than the scatter due to viewing angle for a given orbit/orientation (black crosses).

\subsection{Close Pair Separation Time-Scales}
\begin{figure*}
\includegraphics[width=168mm]{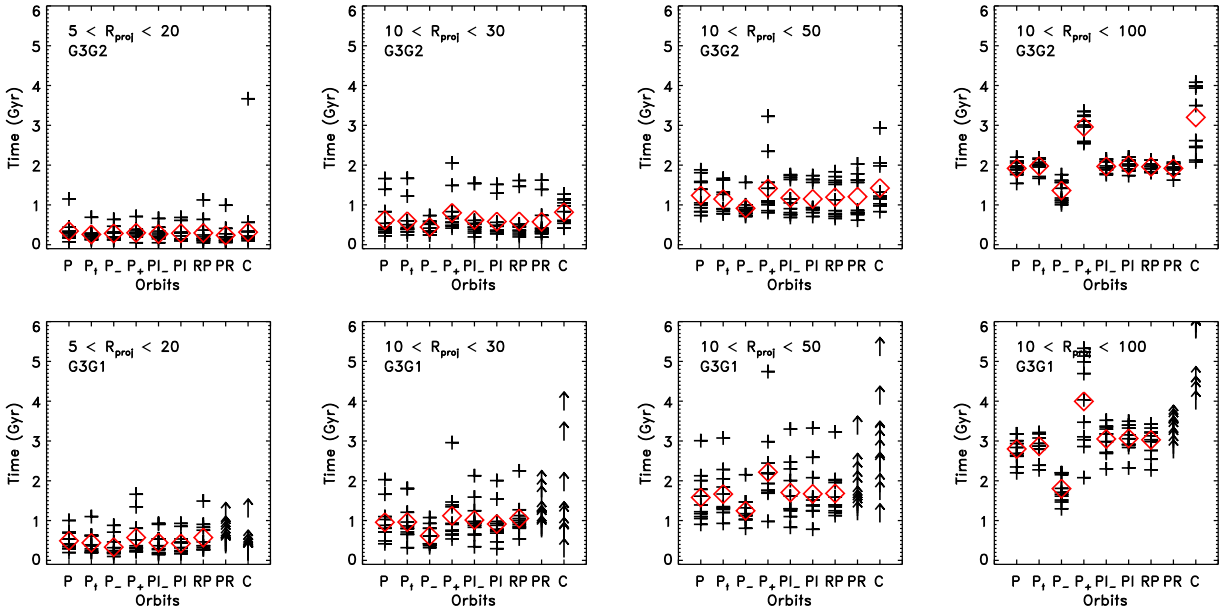}
\caption{Time-scales for projected separations of $5 < R_{proj} < 20$, $10 < R_{proj} < 30$, $10 < R_{proj} < 50$, and $10 < R_{proj} < 100$ $h^{-1}$ kpc
v. orientation/orbits for baryonic mass ratio 3:1 G3G2 mergers (top) and baryonic mass ratio 9:1 G3G1 mergers (bottom).   The time-scales are significantly longer
for circular orbit mergers observed at large projected separations or for high mass ratios.  Symbols and orbits are the same as previous figure. } \label{orbit_p}
\end{figure*}

The time-scales for identifying the simulations as close pairs with projected separations 
$5 < R_{proj} < 20$, $10 < R_{proj} < 30$, $10 < R_{proj} < 50$, and $10 < R_{proj} < 100$ kpc $h^{-1}$ in Table \ref{pairtab} and Figure \ref{orbit_p}.  
As expected from dynamical friction arguments,  the close pair time-scales are
longer for higher mass ratio mergers.  The relative orbits and orientations are less important for 
the  observability time-scales (Figure \ref{orbit_p}),  except when the galaxies are at large separations on circular orbits or start with 
larger pericentric distances. 

We compare our results to the merger time-scales of Boylan-Kolchin, Ma \& Quataert (2008) for GADGET-2 mergers of dark matter haloes, and
of Kitzbichler \& White (2008) derived from the Millenium Simulation semi-analytic model.    The dynamical decay time-scale for the
G3-G2 mergers are $\sim$ 1-2 Gyr shorter than Boylan-Kolchin et al. (2008) dynamical friction time-scale for the dark matter halo merger
of similar total mass, slightly higher total mass ratio (3:1),  initial separation, and similar orbits, possibly because of the effect of baryons. 
The minor G3-G1 merger time-scales for sub-parabolic orbits ($\sim 4$ Gyr) appear to be more consistent with  Boylan-Kolchin et al. (2008) dynamical friction 
time-scale for mergers with similar orbits and mass ratios ($\sim$ 3.5-4.5 Gyr).   

The time-scale for a G3-G2 merger on a circular orbit to appear as a close pair with projected separation $10 < R_{proj} < 50$ kpc $h^{-1}$ (1.2 $\pm$ 0.6 Gyr)
is consistent with Kitzbichler \& White (2008) calculations of the close pair time-scale for a $5 \times 10^{10}$ M$_{\odot}$ stellar mass galaxy pair 
observed with $R_{proj} < 50$ kpc $h^{-1}$ (1.4 Gyr).   The Kitzbichler \& White (2008) close pair time-scales are based on the Croton et al. (2006)
semi-analytic model for galaxies in the dark matter haloes of the Millennium Simulation (Springel et al. 2005),  and assume the 
Binney \& Tremaine (1987) dynamical friction time-scale
for objects on a circular orbit when the subhaloes hosting galaxies can no longer be resolved in the Millennium Simulation.  When compared to our circular
orbit simulations, the majority of parabolic orbits simulations yield  $\sim$ 15-30\% shorter close pair time-scales.  
At larger projected separations ($R_{proj} < 100$ kpc $h^{-1}$),  we find the parabolic orbit simulations have $\sim$ 40\% shorter time-scales than 
the circular orbit simulations.  This implies that the assumption of a
circularized orbit may result in $\sim$ 15-30\% over-estimation of the typical similar mass close pair time-scale at $R_{proj} < 50$ kpc $h^{-1}$, 
and $\sim$ 40\% over-estimation of typical close pair time-scales at $R_{proj} < 100$ kpc $h^{-1}$. 

\begin{figure*}
\includegraphics[width=168mm]{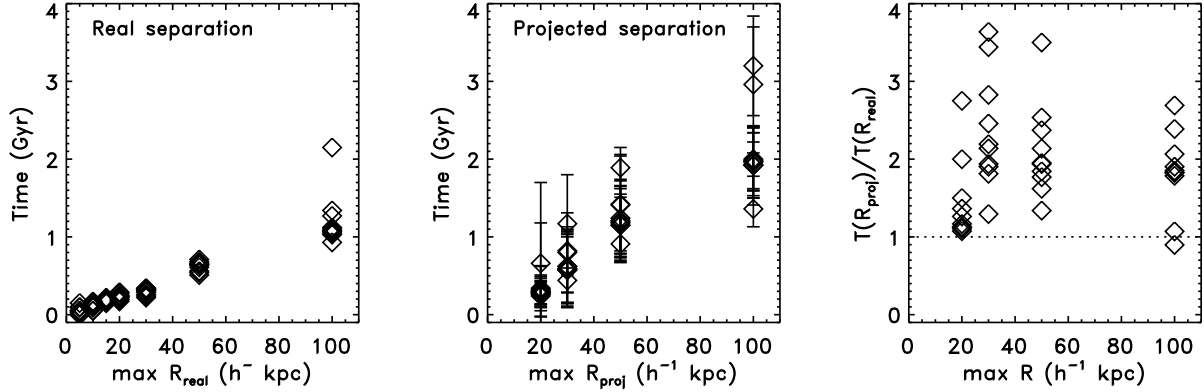}
\caption{{\it Left:} The close pair time-scales calculated for real separations v. real maximum separation for all G3-G2 merger simulations. {\it Center:} 
The average close pair time-scales calculated for projected separation v. projected maximum separation for all G3-G2 merger simulations.  The error-bars
are the standard deviations with viewing angle.  {\it  Right:} The average projected separation time-scale divided by real separation time-scale v. maximum 
separation.   The projected separation time-scales average about twice that of the decay time-scales at that separation in real space because projection
effects make galaxies appear closer than they truly are. } \label{sep}
\end{figure*}

We have compared the time-scales for the true separations of the mergers to those derived for the projected simulations.  
In Figure \ref{sep}, we show that the time-scales for observing a galaxy pair at a given {\it projected} separations are $\sim$ twice as long as
the time-scale for the galaxy pair at the same separation in physical space.  This is because projection effects will always
make galaxy pairs appear closer than they truly are. 


\begin{table*}
  \centering
  \begin{minipage}{168mm}
  \caption{Close Pair Time-Scales}
  \begin{tabular}{@{}lcccc@{}}
\hline
Simulation &  T($5<R_{proj} < 20$)  & T($10 < R_{proj} < 30$)  & T($10 < R_{proj} < 50$)  & T ($10 < R_{proj} < 100$) \\
&   (Gyr)                &   (Gyr)          & (Gyr)  & (Gyr)  \\ 
\hline
\multicolumn{5}{c}{Equal-mass prograde-prograde mergers} \\
\hline    
G3G3P$_t$ &    0.39$\pm$   0.30 &    0.72$\pm$   0.39 &    1.21$\pm$   0.38 &    1.85$\pm$   0.13 \\
G2G2P$_t$\footnote{Time-scale is not calculated when the initial simulation separation is less than maximum projected separation.} &    0.43$\pm$   0.19 &    0.60$\pm$   0.16 &    0.72$\pm$   0.20 &    $-$ \\
G1G1P$_t$$^a$ &    0.59$\pm$   0.12 &    0.52$\pm$   0.16 &    0.64$\pm$   0.19 &    $-$  \\
\hline
\multicolumn{5}{c}{Unequal-mass prograde-prograde mergers} \\
\hline
G3G2P$_t$ &    0.26$\pm$   0.15 &    0.59$\pm$   0.44 &    1.15$\pm$   0.36 &    1.98$\pm$   0.16 \\
G3G1P$_t$ &    0.44$\pm$   0.28 &    0.96$\pm$   0.47 &    1.67$\pm$   0.62 &    2.87$\pm$   0.31 \\
G3G0P$_t$ &   0.58$\pm$   0.42 &    1.23$\pm$   0.53 &   2.75$\pm$   0.54 &   3.44$\pm$   0.34 \\
G2G1P$_t$$^a$ &    0.48$\pm$   0.18 &    0.67$\pm$   0.18 &    0.76$\pm$   0.22 &   $-$ \\  
G2G0P$_t$$^a$ &    1.01$\pm$   0.32 &    1.17$\pm$   0.35 &    1.64$\pm$   0.29 &   $-$ \\ 
G1G0P$_t$$^a$ &    0.79$\pm$   0.32 &    0.78$\pm$   0.31 &    $-$ &    $-$ \\
\hline
\multicolumn{5}{c}{3:1 baryonic mass ratio mergers} \\
\hline
G3G2P      &    0.34$\pm$   0.29 &    0.62$\pm$   0.48 &    1.24$\pm$   0.42 &    1.92$\pm$   0.18 \\
G3G2P$_-$  &    0.28$\pm$   0.15 &    0.44$\pm$   0.16 &    0.91$\pm$   0.23 &    1.36$\pm$   0.28 \\
G3G2P$_+$  &    0.30$\pm$   0.17 &    0.80$\pm$   0.51 &    1.41$\pm$   0.74 &    2.96$\pm$   0.33 \\
G3G2Pl$_t$ &    0.27$\pm$   0.17 &    0.62$\pm$   0.47 &    1.18$\pm$   0.44 &    1.97$\pm$   0.14 \\
G3G2Pl     &    0.29$\pm$   0.20 &    0.58$\pm$   0.42 &    1.15$\pm$   0.40 &    2.00$\pm$   0.15 \\
G3G2RP     &    0.30$\pm$   0.32 &    0.59$\pm$   0.48 &    1.19$\pm$   0.46 &    1.96$\pm$   0.11 \\
G3G2PR     &    0.24$\pm$   0.27 &    0.57$\pm$   0.48 &    1.21$\pm$   0.51 &    1.92$\pm$   0.13 \\
G3G2C      &    0.32$\pm$   1.38 &    0.82$\pm$   0.31 &    1.42$\pm$   0.64 &    3.20$\pm$   0.84 \\
\hline
\multicolumn{5}{c}{9:1 baryonic mass ratio mergers} \\
\hline
G3G1P       &    0.49$\pm$   0.24 &    0.96$\pm$   0.49 &    1.58$\pm$   0.62 &    2.80$\pm$   0.34\\
G3G1P$_-$   &    0.33$\pm$   0.25 &    0.62$\pm$   0.26 &    1.25$\pm$   0.36 &    1.80$\pm$   0.33\\
G3G1P$_+$   &    0.57$\pm$   0.49 &    1.12$\pm$   0.69 &    2.22$\pm$   0.98 &    4.00$\pm$   1.14\\
G3G1Pl$_t$  &    0.44$\pm$   0.28 &    1.01$\pm$   0.49 &    1.70$\pm$   0.74 &    3.05$\pm$   0.36\\
G3G1Pl      &    0.42$\pm$   0.26 &    0.91$\pm$   0.48 &    1.67$\pm$   0.73 &    3.06$\pm$   0.34\\
G3G1RP      &    0.57$\pm$   0.39 &    1.05$\pm$   0.44 &    1.68$\pm$   0.61 &    3.03$\pm$   0.37\\
G3G1PR\footnote{The simulation does not merge within 6 Gyr.}     & $\ge$    0.43$\pm$   0.25 &   $\ge$  1.01$\pm$   0.41 &  $\ge$   1.64$\pm$   0.63 &  $\ge$   3.04$\pm$   0.27\\
G3G1C$^b$       & $\ge$    0.18$\pm$   0.31 &   $\ge$  1.23$\pm$   1.16 &  $\ge$   2.64$\pm$   1.16 &  $\ge$   5.06$\pm$   0.95\\
\hline
\end{tabular}\label{pairtab}
\end{minipage}
\end{table*}

\section{Summary and Implications}
We have analysed the $g$-band quantitative morphologies and projected pair separations for a large suite of {\sc GADGET/SUNRISE} 
simulations of unequal-mass disc galaxy mergers.  These merger simulations span a range in baryonic mass ratio from 1:1 to 39:1, and include
a range of orientations and orbital parameters for the merging galaxies.   The initial galaxies have baryonic gas fractions matched to local disc galaxies. 
We determine the observability time-scales for identifying these simulated mergers using the quantitative morphology classification in 
$G-M_{20}$, $G-A$, and $A$ and as close pairs with projected separations $5 < R_{proj} < 20$, $10 < R_{proj} < 30$, $10 < R_{proj} < 50$, 
and $10 < R_{proj} < 100$ kpc $h^{-1}$.  Our main conclusions are as follows:

$\bullet$ Different morphological approaches identify mergers with different mass ratios.  $G-M_{20}$ is 
sensitive to mergers with baryonic mass ratios between 1:1 and at least 9:1,  and does not show a correlation of observability time-scale 
with baryonic mass ratio less than 9:1.  Typical $G-M_{20}$ time-scales are $0.2-0.4$ Gyr. 
Asymmetry, on the other hand, is much less likely to find minor mergers with baryonic mass ratios greater than 4:1 when 
gas-fractions typical of local galaxies are assumed.  The asymmetry time-scales for major mergers of moderate gas-fraction discs are also $0.2-0.4$ Gyr, and
less than 0.06 Gyr for minor mergers of moderate gas-fraction.  Hence, the $G-M_{20}$ method detects both major and minor mergers, while the majority of 
asymmetric objects in the local universe will be major mergers.   From visual inspection of the simulation images, 
we find that $G-M_{20}$ detects mergers when double nuclei are enclosed in a common envelope.  If the nucleus of the merging 
satellite galaxy is among the brightest 20\% of the pixels, it will produce high $M_{20}$ values at close passages and final merger.  $A$ is more sensitive
to the relative brightnesses of the nuclei, and is strongest when the merging nuclei are of similar luminosity. 
 
$\bullet$ The orbits and orientations have little effect on the overall time-scale for strong morphological disturbances, but do effect the
time-scale for identifing galaxy pairs.
The time-scales for close pairs with projected separations $R_{proj} < 50$ kpc $h^{-1}$ are $\sim$ 15$-$30\% longer for circular orbits than
parabolic orbits.   For close pairs with projected separations $\geq 100$ kpc $h^{-1}$, 
we find that circular orbits result in close pair time-scales $\sim$ 40\% greater than those found for mergers initially on
parabolic orbits. 

$\bullet$ Our major merger remnants observed $>$ 1 Gyr after the final merger have quantitative morphologies and star-formation rates more like early-type spirals
than elliptical/S0 galaxies.   We note that only supernova feedback is including in our simulations,  and other quenching mechanisms (e.g. AGN feedback) 
may prevent the reformation of a disc and additional star-formation. We find that minor mergers have a minimal effect
on the $G$, $M_{20}$, $C$, and $A$ values of their remnants. 

In Paper 1, we found that the various methods for finding galaxy mergers identify a given equal-mass merger at different stages in the merger process and
for varying periods of time.  However, these different time-scales for major mergers are not sufficient to explain the observed differences between
the fractions of galaxies in close pairs and the fraction of galaxies with visually or quantitatively disturbed morphologies (e.g. Lotz et al. 2008a, Jogee et al. 2009).   
With this work, we now find that the two main quantitative morphology approaches -- $G-M_{20}$ and $A$ -- are sensitive to different ranges in merger mass ratios.   

The range of merger mass ratio detected by the different techniques has important implications for computing the galaxy merger rate. 
In a given galaxy survey,  the galaxy merger candidates identified by $G-M_{20}$ contain both minor mergers and major majors,  
while $A$ is more likely to find major mergers (assuming local disc galaxy gas fractions).  
Because $G-M_{20}$ identifies the merger stages when galaxies exhibit double nuclei, these minor mergers may also be found by visual classification.   
Close kinematic pair studies typically limit the mass or flux ratios of the paired galaxies
to $\leq$ 2:1 $-$ 4:1, and hence select major merger events.  Thus our results here may resolve many of the discrepancies in the merger fractions and
merger rates found using different methods.    

Finally, we note that we have not explored the properties of mixed-morphology mergers (e.g. disc-spheroid: Johansson et al. 2009) or 
spheriod-spheroid mergers (e.g. Naab, Khochfar, \& Burkert 2006; Bell et al. 2006; Boylan-Kolchin, Ma, \& Quataert 2005), 
which are likely to give shorter morphological disturbance time-scales.   We expect the time-scales derived in this paper to be upper
limits to the morphological disturbance time-scales of gas-poor mergers. 
We explore the effect of gas fraction on the morphologies of merging discs in a companion paper to this work (Paper 3),  and  find that 
like mass ratio, gas-fraction is an important parameter in the morphological disturbance time-scales and the calculation of the galaxy merger rate.

This series of papers are the needed first steps for understanding which parameters are important for the time-scales of morphological
disturbances, how these relate to close pair time-scales,  and for deriving a merger rate from the observations of morphologically disturbed galaxies.
Although the parameter space explored in these works is not exhaustive,  we now understand which parameters are most important for
the duration of morphological disturbances  $-$ mass ratio and gas-fraction.  We will apply these merger observability time-scales to observations of the merger
fraction using $G-M_{20}$, $A$, and close pairs in a future paper.  By comparing the
minor $+$ major merger frequency detected by $G-M_{20}$ to the major merger frequency detected by either $A$ or close pairs, we will be able to
place some of the first quantitative constraints on the minor merger rate. 

JML acknowledges support from the NOAO Leo Goldberg Fellowship, NASA grants NAG5-11513 and HST-AR-9998, and 
would like to thank P. Madau for support during this project. 
PJ was supported by programs HST-AR-10678, HST-AR-10958, and HST-AR-11958, provided by NASA through grants from the Space Telescope Science
Institute, which is operated by the Association of Universities for Research in Astronomy, Incorporated, 
under NASA contract NAS5-26555, and  by the Spitzer Space Telescope Theoretical Research Program, through a contract 
issued by the Jet Propulsion Laboratory, California Institute of Technology under a contract with NASA.
TJC was supported by a grant from the W.M. Keck Foundation. JRP's research was supported by NASA ATP grant NNX07AG94G.

This research used computational resources of the
NASA Advanced Supercomputing Division (NAS) and the National Energy
Research Scientific Computing Center (NERSC), which is supported by
the Office of Science of the U.S. Department of Energy.

\end{document}